\documentclass[12pt,english,journal=mamobx,manuscript=article,maxauthors=15,biblabel=plain]{revtex4-1}
\usepackage[T1]{fontenc}
\usepackage[latin9]{inputenc}
\usepackage[letterpaper]{geometry}
\usepackage{amsthm}
\usepackage{amsmath}
\usepackage{amssymb}
\usepackage{graphicx}

\makeatletter

\providecommand{\tabularnewline}{\\}

\numberwithin{equation}{section}
\numberwithin{figure}{section}

\makeatother

\usepackage{babel}
\begin{document}

\title{Arm retraction dynamics in dense polymer brushes}

\author{M. Lang, M. Werner, R. Dockhorn, and T. Kreer}
\begin{abstract}
Large scale Monte Carlo simulations of dense layers of grafted polymer
chains in good solvent conditions are used to explore the relaxation
of a polymer brush. Monomer displacements are analyzed for the directions
parallel and perpendicular to the grafting plane. Auto-correlation
functions of individual segments or chain sections are monitored as
function of time. We demonstrate that the terminal relaxation time
$\tau$ of grafted layers well in the brush regime grows exponentially
with degree of polymerization $N$ of the chains, $\tau\propto N^{3}\exp(N/N_{e})$,
with $N_{e}$ the entanglement degree of polymerization in the brush.
One specific feature of entangled polymer brushes is that the late
time relaxation of the perpendicular component coincides for all segments.
We use this observation to extract the terminal relaxation time of
an entangled brush.
\end{abstract}
\maketitle

\section{Introduction}

Polymer brushes are layers of polymer chains grafted densely onto
a surface. These brushes can be used in various applications ranging
from drug delivery \cite{Torchilin}, colloid stabilization \cite{Napper,Russel},
and lubrication \cite{Klein1,Klein2,Moru,Galuschko,Spirin}, to switchable
amphiphilic surfaces \cite{Uhlmann}. In recent years, brushes made
of more complex architectures like stars \cite{Cui}, cyclic polymers
\cite{He-1}, dendrimers \cite{Cui-1} along with cross-linked brushes
\cite{Hoffmann,Lang Hoffmann}, or systems containing nano-inclusions
\cite{Spirin-1,Merlitz} were analyzed.

The relaxation dynamics of polymer brushes is of great importance
for understanding the polymer-brush lubrication \cite{Galuschko,Kreer-3},
which itself is relevant for medical implants \cite{Klein,Moru,Zdyrko},
drug delivery \cite{Goicochea}, microfluidic devices \cite{Advincula,Neratova},
and so forth. Also, the ultra-low friction of human joints motivated
work on out-of-equilibrium conformations and dynamics of polymer brush
bi-layers \cite{Spirin-1,Binder,Spirin-2,Kreer} and kinetic barriers
have been identified as limiting factors for the formation of nano-aggregates
using bridging of surfaces via polymer brushes \cite{Balko} to provide
just some examples of current research.

With these increasingly diverse research topics and applications,
a general understanding of the most fundamental case, a densely grafted
layer of monodisperse linear chains, is highly in demand. Even though
a large number of self-consistent field approaches have been published
previously, see for instance Refs. \cite{Semenov,MWC,Skvortsov,Amoskov,Netz},
a self-consistent model that leads to the same free energy as scaling
models \cite{Alexander,De Gennes} was developed only recently \cite{Romeis}.
Concerning the relaxation dynamics, simulation data and theoretical
models still are controversial as briefly summarized below.

For a polymer brush, it is generally assumed that the grafting density
$\sigma$ defines the correlation length, 
\begin{equation}
\xi\approx\sigma^{-1/2},\label{eq:correlation}
\end{equation}
at which neighboring strands start to interact \cite{Alexander,De Gennes}.
This length is spanned in average by a short strands of 
\begin{equation}
g\approx(\xi/b)^{1/\nu}\approx\sigma^{-1/(2\nu)}b^{-1/\nu}\label{eq:g}
\end{equation}
Kuhn segments. Here, $b$ is the root mean square size of a Kuhn segment
and $\nu\approx0.588$ is the Flory exponent describing polymer conformations
in good solvent \cite{LeGuillou,Baker}, while $\nu=1/2$ for $\theta$
solvents. In the ``brush regime'', the overlap of the chains leads
to a swelling of polymers made of $N$ Kuhn segments to a layer of
thickness \cite{Alexander,De Gennes} 
\begin{equation}
H\approx\xi N/g\approx N\sigma^{(1-\nu)/(2\nu)}b^{1/\nu}.\label{eq:H}
\end{equation}
For friction dominated dynamics, the relaxation time of a correlation
volume, $\tau_{\xi}$, can be estimated \cite{RubinsteinColby} as
\begin{equation}
\tau_{\xi}\approx\tau_{0}g^{2\nu+1}\approx\tau_{0}\sigma^{-(2\nu+1)/(2\nu)}b^{-(2\nu+1)/\nu}\label{eq:taux}
\end{equation}
using the monomer relaxation time, $\tau_{0}$. For non-entangled
brushes, the terminal relaxation time was first estimated in Ref.
\cite{Klushin} using a dumbbell model. The stretched chain conformations
in direction perpendicular to the grafting plane lead to an extended
relaxation time 
\begin{equation}
\tau_{\perp,d}\propto\tau_{\xi}(N/g)^{3}\propto\sigma^{(1-\nu)/\nu}b^{-(2-2\nu)/\nu}N^{3},\label{eq:tauperp}
\end{equation}
as compared to the parallel directions, where the relaxation time
scales as 
\begin{equation}
\tau_{\parallel,d}\propto\tau_{\xi}(N/g)^{2}\propto\sigma^{-(2\nu-1)/(2\nu)}b^{-(2\nu-1)/\nu}N^{2}.\label{eq:tauparallel}
\end{equation}
In good solvent, both results are approximately $\propto\sigma^{2/3}$
and $\sigma^{-1/6}$ respectively, while these results become $\propto\sigma^{1}$
and $\propto\sigma^{0}$ for $\theta$ solvents. Note that a similar
result would be obtained using the Rouse model as basis for computation.

Hydrodynamics up to the scale of a blob renormalizes the blob relaxation
time $\tau_{\xi}\approx\tau_{0}g^{3\nu}$ for good solvents as introduced
first in Ref. \cite{Johner} for a polymer brush. This leads to relaxation
times for this model with 
\begin{equation}
\tau_{\perp,h}\propto\sigma^{3(1-\nu)/(2\nu)}N^{3}\approx\sigma^{1.05}N^{3}\label{eq:tauperpJ}
\end{equation}
in perpendicular direction and 
\begin{equation}
\tau_{\parallel,h}\propto\sigma^{(2-3\nu)/(2\nu)}N^{2}\approx\sigma^{0.20}N^{2}\label{eq:tauparJ}
\end{equation}
in parallel direction ($\theta$-solvent: $\propto\sigma^{3/2}$ and
$\propto\sigma^{1/2}$, respectively). The difference to the friction
dominated case above lies only in the modified scaling as function
of the grafting density $\sigma$. Note also that the stretching field
of the brush allows in both cases for extra large dynamic fluctuations
of the chain extensions $\propto N$ in perpendicular direction \cite{Klushin}.
Non-entangled chains in a brush, therefore, are ``swollen'' coils
and relax in self-similar manner on all subscales in a first approximation
(up to small corrections with respect to the particular shape of the
brush potential).

A qualitatively different result is obtained for entangled polymer
brushes. Here, the terminal relaxation time is expected \cite{Witten}
to increase exponentially as function of the number of effective monomers
(here: $N/g$ correlation volumes) for large $N$, 
\begin{equation}
\tau\propto\left(N/g\right)^{3}\exp(N/g),\label{eq:relax}
\end{equation}
since terminal relaxation is governed by the arm retraction dynamics
of the grafted polymers similar to the relaxation of branched polymers.

Let us number the entangled sections starting from the grafting point
with a variable $s$ up to $s_{max}=N/N_{e}$ at the free end. Here,
$N_{e}$ is the entanglement degree of polymerization. For arm retraction
dynamics, atypically large tube length fluctuations are required to
release entanglements of sections $s\lesssim s_{max}-\left(N/N_{e}\right)^{1/2}$.
Usually, this is modeled as a thermally activated process in an effective
potential that is approximated by a parabola \cite{RubinsteinColby}
\begin{equation}
U(s)\approx\frac{\gamma kT}{2}\frac{(s_{max}-s)^{2}N_{e}}{N},\label{eq:U(s)}
\end{equation}
where $\gamma$ is an effective dimensionless spring constant of order
unity, $k$ the Boltzmann constant and $T$ the absolute temperature.
Note that because of this effective potential we have to consider
now ``stretched'' chains with reduced dynamic fluctuations instead
of ``swollen'' ones. This approximation leads to an exponential
decrease $\propto\exp\left(-\frac{\gamma(s_{max}-s)^{2}N_{e}}{2N}\right)$
of the probability for finding tube lengths that are reduced to $s$
sections. As consequence, the time it takes to release entanglement
number $s$ by retraction grows exponentially for decreasing $s$
\begin{equation}
t\left(s\right)\propto\tau_{a}\exp\left(\gamma(s_{max}-s)^{2}N_{e}/(2N)\right).\label{eq:t(s)}
\end{equation}
$\tau_{a}$ is the ``attempt time'' of retractive motion and related
to the Rouse time of the polymer in the particular environment. Due
to the exponential increase in relaxation times, the number $(s_{max}-s)$
of entanglements released per chain is roughly the same for all chains
at time $t$ and grows logarithmically at large times: 
\begin{equation}
s_{max}-s(t)\propto(2N/(\gamma N_{e}))^{1/2}\left(\ln(t/\tau_{a})\right)^{1/2}.\label{eq:s(t)}
\end{equation}

O'Connor and McLeish used arm retraction dynamics to study the adhesion
of a polymer network on a dry brush \cite{OConnor,OConnor2}. They
argue that for a brush penetrating a network, the brush self-consistent
potential should enter only as a small (logarithmic) correction to
$\tau_{a}$, while the exponential increase with the degree of polymerization
$N$ is not modified. We expect a similar behaviour for monodisperse
brushes in good solvent. However, the different Rouse times in parallel
and perpendicular direction, see equations (\ref{eq:tauperp}) and
(\ref{eq:tauparallel}), lead to a shorter attempt time $\tau_{a}$
for retractive motion in parallel directions. This could drive disentanglement
of the chains in parallel directions without the need for retractive
motions in perpendicular direction.

It is also instructive to mention the relaxation of a single chain
in an array of straight rods that are grafted perpendicular to a surface
as a limiting case of relaxation in anisotropic entanglements. Here,
the retraction dynamics in directions parallel to the grafting plane
becomes decoupled from the Rouse relaxation in perpendicular direction
on time scales beyond the entanglement time. A similar behavior as
for this limiting case might be visible in an entangled polymer brush
at intermediate times between the entanglement relaxation times in
parallel and perpendicular direction, $\tau_{e\parallel}<t<\tau_{e\perp}$,
if $\tau_{e\perp}/\tau_{e\parallel}\approx N_{e}/g$ is sufficiently
large. Note that $\tau_{e\parallel}$ and $\tau_{e\perp}$ are readily
obtained from inserting $N_{e}$ instead of $N$ into equations (\ref{eq:tauperp})
and (\ref{eq:tauparallel}) or, if hydrodynamics is important, equations
(\ref{eq:tauperpJ}) and (\ref{eq:tauparJ}).

Finally, we have to mention that entangled brushes refer to a case,
where the chains inside the confining tubes are largely ``over-stretched''.
This point follows from $N_{e}>g$ in a semi-dilute solution of flexible
polymers \cite{RubinsteinColby} and the fact that the correlation
blob size equals the tension blob size in a brush \cite{De Gennes}.
Thus, the tension blobs are smaller than the length scale that is
typically associated with an entanglement (``tube diameter''). This
leads to entangled sections of size $a_{\parallel}\approx\xi(N_{e}/g)^{1/2}$
and $a_{\perp}\approx\xi(N_{e}/g)$ that consist of a small series
\cite{footnote2} of essentially pairwise chain contacts that are
pulled taught within volumes of order $\xi^{3}$. This is quite different
to entanglements in a melt with a lateral tube confinement of order
$a\approx bN_{e}^{1/2}$. In consequence, small chain sections of
extra $g$ monomers that can fold a distance $\xi$ away from the
primitive path of this ``over-stretched'' tube require $kT$ on
energy and, thus, become exponentially rare. Hence, we expect to determine
the blob size $\xi$ instead of the size of an entangled section when
analyzing the lateral confinement along the tube in the vicinity of
the grafting plane.

The exponentially increasing relaxation times mentioned above were
successfully tested in several experiments \cite{Bureau,Geoghegan,Clarke,Chenneviere}
on the adhesion of bulk polymer materials like networks or entangled
melts on a polymer brush. Previous simulation works \cite{Murat,Lai,Lai-1,Marko,Binder-1,Grest,He,Reith},
however, confirm mainly the estimates for the non-entangled case,
see Table \ref{tab:Scaling-of-terminal}. Only the authors of one
work \cite{Marko} suggest an exponential increase of the relaxation
time already for a rather small number of $N/g$. In a simulation
of two brushes sliding on top of each other, also an almost logarithmic
relaxation of the end-to-end vector of the chains was found for rather
small $N=60$ and a grafting density only about twice as high as the
overlap grafting density \cite{Kreer-2}. More recent work \cite{Reith}
covering a much larger $N/g$ observes an effective relaxation time
$\propto N^{3.7}$ in disagreement to previous theoretical works \cite{Witten,Klushin,Johner,OConnor,OConnor2,Rubinstein-1}.
Quite unexpectedly, the largest relaxation times were found near the
middle of the chains, where the relaxation in parallel directions
was slower than in perpendicular direction. A qualitatively similar
result was obtained in a second study for spherical brushes \cite{Verso}.

\begin{table}
\begin{tabular}{|c|c|c|c|c|c|c|c|}
\hline 
Method & System & direction & analysis & $N$ & $\sigma$ & $\tau$ & Ref.\tabularnewline
\hline 
\hline 
MD & a-thermal & O & EAT & $10\le100$ & $0.03\le0.1$ & $\propto N^{2.5\pm0.1}\sigma^{0.5}$ & \cite{Murat}\tabularnewline
\hline 
BFM & a-thermal & $\perp$ & CAT & $10\le60$ & $0.025\le0.2$ & $\propto N^{3.0\pm0.01}\sigma^{0.83\pm0.08}$ & \cite{Lai}\tabularnewline
\hline 
BFM & $\Theta$-solvent & $\perp$ & CAT & $20\le50$ & $0.05\le0.125$ & $\propto N^{3}\sigma^{1.6}$ & \cite{Lai-1}\tabularnewline
\hline 
LMC & a-thermal & O & DS & $50\le100$ & $0.02\le0.08$ & $\propto N^{2}\sigma^{-1/6}$ & \cite{Marko}\tabularnewline
\hline 
LMC & a-thermal & $\perp$ & DS & $50\le100$ & $0.02\le0.08$ & $\propto N^{3}\sigma^{2/3}$ & \cite{Marko}\tabularnewline
\hline 
LMC & a-thermal & $\perp$ \& O & DS & $50$ & $0.12$ & exponential? & \cite{Marko}\tabularnewline
\hline 
BFM & a-thermal & $\parallel$ & CAT & $20\le60$ & $0.05\le0.1$ & $\propto N^{2}\sigma^{-1/6}$ & \cite{Binder-1}\tabularnewline
\hline 
BFM & a-thermal & $\perp$ & CAT & $10\le80$ & $0.025\le0.2$ & $\propto N^{3}\sigma^{2/3}$ & \cite{Binder-1}\tabularnewline
\hline 
MD & brush in melt & O & EAT, RAT & $50\le200$ & $0.01\le0.41$ & $\propto N^{3}\sigma$ & \cite{Grest}\tabularnewline
\hline 
MD & a-thermal & $\parallel$ & MSD & $64\le128$ & $0.013\le0.492$ & $\propto N^{2}\sigma^{-1/6}$ & \cite{He}\tabularnewline
\hline 
MD & a-thermal & $\perp$ & MSD & $64\le128$ & $0.013\le0.492$ & $\propto N^{3}$ & \cite{He}\tabularnewline
\hline 
MC+MD & a-thermal & $\perp$ & MSD, RAT & $16\le64$ & $0.125$ & $\propto N^{3}$ & \cite{Reith}\tabularnewline
\hline 
MD & a-thermal & $\parallel$ \& $\perp$ & MSD, RAT & $64\le512$ & $0.125$ & $\propto N^{3.7}$ & \cite{Reith}\tabularnewline
\hline 
BFM & a-thermal & $\parallel$, $\perp$ \& O & MSD, TTAA & $16\le128$ & $1/16\le1/4$ & $\propto N^{3}\exp(N)$ & this work\tabularnewline
\hline 
\end{tabular}

\caption{\label{tab:Scaling-of-terminal}Terminal relaxation times, as determined
by computer simulations. We compare results from molecular dynamics
simulations (MD), the Bond Fluctuation Method (BFM), off-lattice Monte
Carlo (MC), and lattice Monte Carlo simulations (LMC). All ``systems''
concern homopolymer brushes of monodisperse chain lengths under different
solvent conditions. The analyzed directions are either isotropic (``O''),
perpendicular to the grafting plane (``$\perp$''), or parallel
to the grafting plane (``$\parallel$''). EAT, CAT, RAT, and DS
are the end-to-end vector auto-correlation time, the directional component
auto-correlation time, radius of gyration (or its spacial components)
auto-correlation time, or dynamic scattering, respectively. TTAA denotes
the terminal time auto-correlation analysis as introduced in section
\ref{sub:Auto-correlation-functions}. For EAT, CAT, RAT, and DS,
the relaxation time was always defined by the decay of the function
to $1/e$. MSD refers to an analysis of the mean square displacements.
The data range for the degree of polymerization, $N$, and the grafting
density, $\sigma$, is indicated along with the suggested scaling
of the terminal relaxation time, $\tau$.}
\end{table}

We attempt to resolve this puzzle by a comparison of different observables
(monomer mean square displacements and auto-correlation functions)
that are analyzed such that the differences concerning entangled and
non-entangled brushes or concerning other points raised in the introduction
stand out. In addition, we test the concepts introduced above and
discuss how to correctly determine the terminal relaxation time of
an entangled polymer brush.

\section{Computer Simulations}

Since we want to understand the universal behavior of brushes formed
from flexible chains, we use the Bond Fluctuation Model (BFM) as introduced
by Carmesin and Kremer \cite{Carmesin} and extended to three dimensions
by Deutsch and Binder \cite{Deutsch}. In this model, each monomer
is represented by a cube of eight lattice positions on a simple cubic
lattice. The monomers of a chain are connected by bond vectors out
of a particular vector set. The motion of the monomers occurs via
random displacements by one lattice unit along a lattice axis, where
those moves are rejected for which the new particle position leads
to overlap of monomers or to bond vectors not contained in the set.
Solvent is treated implicitly in the a-thermal limit. Because of the
neglect of momentum and solvent, no hydrodynamic effects are modeled
and dynamics is friction dominated. In order to collect data over
an extremely long time interval, we use a GPU implementation of the
BFM \cite{Nedelcu}, which leads to equivalent data concerning the
dynamics as compared to the original implementations at a slightly
enlarged effective excluded volume.

For the present study, we focus on brushes at rather large grafting
densities and chain lengths in order to explore the dynamics of entangled
brushes. Periodic boundary conditions were applied in the directions
parallel to the grafting plane. The perpedicular dimension of the
simulation box was chosen sufficiently large to avoid confinement
by the box size.

\begin{table}
\begin{tabular}{|c|c|c|c|c|c|c|c|c|}
\hline 
Sample & \#1 & \#2 & \#3 & \#4 & \#5 & \#6 & \#7 & \#8\tabularnewline
\hline 
\hline 
$N$ & 16 & 32 & 64 & 64 & 64 & 128 & 128 & 128\tabularnewline
\hline 
$\sigma$ & $1/4$ & $1/4$ & $1/4$ & $"1/9"$ & $1/16$ & $1/4$ & $"1/9"$ & $1/16$\tabularnewline
\hline 
$M$ & 1024 & 1024 & 1024 & 1764 & 4096 & 1024 & 882 & 1024\tabularnewline
\hline 
$\tau$ {[}MCS{]} & $2.5\cdot10^{5}$ & $2.4\cdot10^{6}$ & $4.9\cdot10^{7}$ & $1.0\cdot10^{7}$ & $5.3\cdot10^{6}$ & $8.6\cdot10^{9}$ & $4.9\cdot10^{8}$ & $1.0\cdot10^{8}$\tabularnewline
\hline 
$t_{tot}$ {[}MCS{]} & $1.2\cdot10^{7}$ & $7.7\cdot10^{9}$ & $2.2\cdot10^{10}$ & $5\cdot10^{8}$ & $2.9\cdot10^{9}$ & $2.7\cdot10^{10}$ & $9.9\cdot10^{9}$ & $7.6\cdot10^{9}$\tabularnewline
\hline 
\end{tabular}

\caption{\label{tab:Simulation-parameters-for}Simulation parameters for degree
of polymerization, $N$, grafting density, $\sigma$, and number of
grafted chains per sample, $M$. Note that $\sigma$ is normalized
to the maximum possible grafting density in the BFM model, which is
0.25 in square lattice units. The exact grafting density for samples
4 and 7 is $441/4096$ (we use ``$1/9$'' to simplify notation in
Tables and Figures but use the exact value for computation). The terminal
relaxation time $\tau$ is determined using the perpendicular component
of the position vector auto-correlation function as described in section
\ref{sub:Auto-correlation-functions}. The last row indicates the
total duration $t_{tot}$ of the simulation runs.}
\end{table}

Initial chain conformations were relaxed prior to analysis for several
terminal relaxation times $\tau$, which was determined as described
below in section \ref{sec:Results}. Production runs spanned at least
20 relaxation times. The only exception is sample \#6, for which only
3 relaxation times could be realized, see Table \ref{tab:Simulation-parameters-for}.
Since no qualitative variation of the results of sample \#6 (except
of a clearly larger scatter of the data) was detected, we consider
all samples as sufficiently equilibrated for analysis.

\section{\label{sec:Results}Results}

In this section, we summarize only the results for the dynamics of
polymer brushes, since the data on chain conformations (with extra
data in the low $\sigma$ limit) were published previously in Ref.
\cite{Romeis}. Let $n$ denote the number of the monomers along a
chain as counted from the grafting point. $\mathbf{R}_{n}(t)$ is
the position vector of monomer $n$ at time $t$ with respect to the
grafting point of the chain. We consider long chains of $N$ monomers
in the brush regime with $N\gg g$, where $g$ is the number of monomers
per concentration blob. $b$ is the ensemble average root mean square
length of a bond between any two connected monomers. Ensemble averages
over all chains are denoted by angular brackets $\left\langle ...\right\rangle $.
The statistics of all dynamic quantities is improved by averaging
over a large set of different starting times. Local dynamics are analyzed
by means of the normalized bond auto-correlation function 
\begin{equation}
b_{c}(n,\Delta t)=\frac{\left\langle \mathbf{b}_{n}(t)-\left\langle \mathbf{b}_{n}\right\rangle ,\mathbf{b}_{n}(t+\Delta t)-\left\langle \mathbf{b}_{n}\right\rangle \right\rangle }{\left\langle \left(\mathbf{b}_{n}(t)-\left\langle \mathbf{b}_{n}\right\rangle \right)^{2}\right\rangle }\label{eq:bc}
\end{equation}
for oriented chains, whereas relaxation from the free end is analyzed
by using the auto-correlation function
\begin{equation}
a_{c}(n,\Delta t)=\frac{\left\langle \mathbf{R}_{n}(t)-\left\langle \mathbf{R}_{n}\right\rangle ,\mathbf{R}_{n}(t+\Delta t)-\left\langle \mathbf{R}_{n}\right\rangle \right\rangle }{\left\langle \left(\mathbf{R}_{n}(t)-\left\langle \mathbf{R}_{n}\right\rangle \right)^{2}\right\rangle },\label{eq:ac}
\end{equation}
of the position vector $\mathbf{R}_{n}(t)$. Here, $b_{n}$ is the
root mean square length of the bond vector $\mathbf{b}_{n}=\mathbf{R}_{n}-\mathbf{R}_{n-1}$
between monomers $n$ and $n-1$. $\mathbf{R}_{n-1}$ is the zero
vector for $n=1$. The average vectors $\left\langle \mathbf{x}_{n}\right\rangle ,$
with $\mathbf{x}=\mathbf{R},\mathbf{b}$ are required for proper normalization
of the directional contributions perpendicular to the grafting plane.
The time average vector components in parallel directions are assumed
to be zero and the above expressions are equivalent to standard normalized
vector auto-correlation functions in this case. As in the introduction,
the directional contributions parallel and perpendicular to the grafting
plane are denoted by the corresponding subscripts $||$ and $\perp$.
Directional mean square displacements (or monomer relaxation times,
auto-correlation times, etc ...) are determined using the directional
components of vectors $\mathbf{b}_{n}$ or $\mathbf{R}_{n}$, respectively.

\subsection{Monomer mean square displacements}

The mean square displacement of monomer $n$ can be written as a function
of the time interval $\Delta t$ for an arbitrary starting time $t$
\begin{equation}
\left\langle \left[d_{n}(\Delta t)\right]^{2}\right\rangle =\left\langle \left[\mathbf{R}_{n}(t+\Delta t)-\mathbf{R}_{n}(t)\right]^{2}\right\rangle .\label{eq:msd}
\end{equation}

\begin{figure}
\includegraphics[width=1\columnwidth]{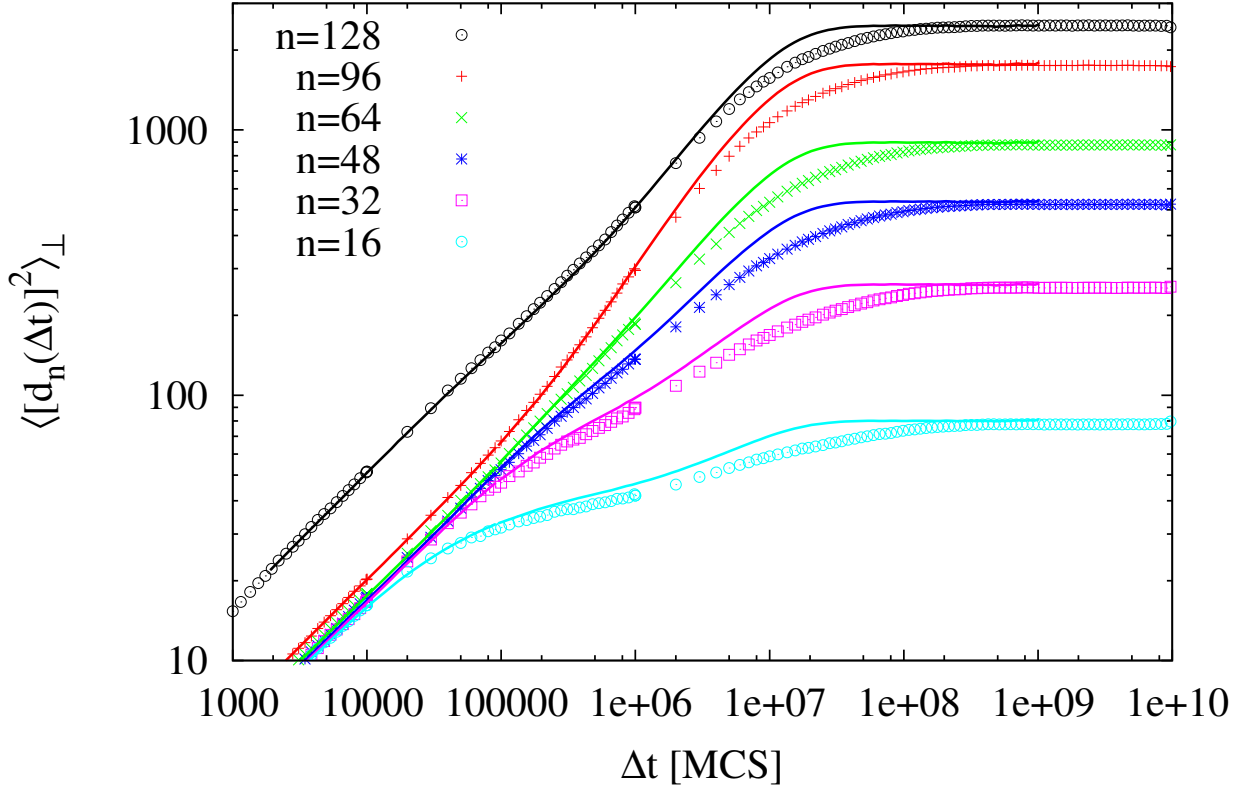}

\caption{\label{fig:Monomer-mean-square}Monomer mean square displacements
in perpendicular direction, $\left\langle \left[d_{n}(\Delta t)\right]^{2}\right\rangle _{\perp}$,
for different monomers $n$ of sample \#7 with $N=128$ and $\sigma=1/9$.
The lines show the numerical solution of the Rouse model \cite{Reith,Hoffmann-1}
with mean square displacements $\propto t^{1/2}$ that was rescaled
to match maximum displacements and local friction coefficients, see
text for details.}
\end{figure}

\begin{figure}
\includegraphics[width=1\columnwidth]{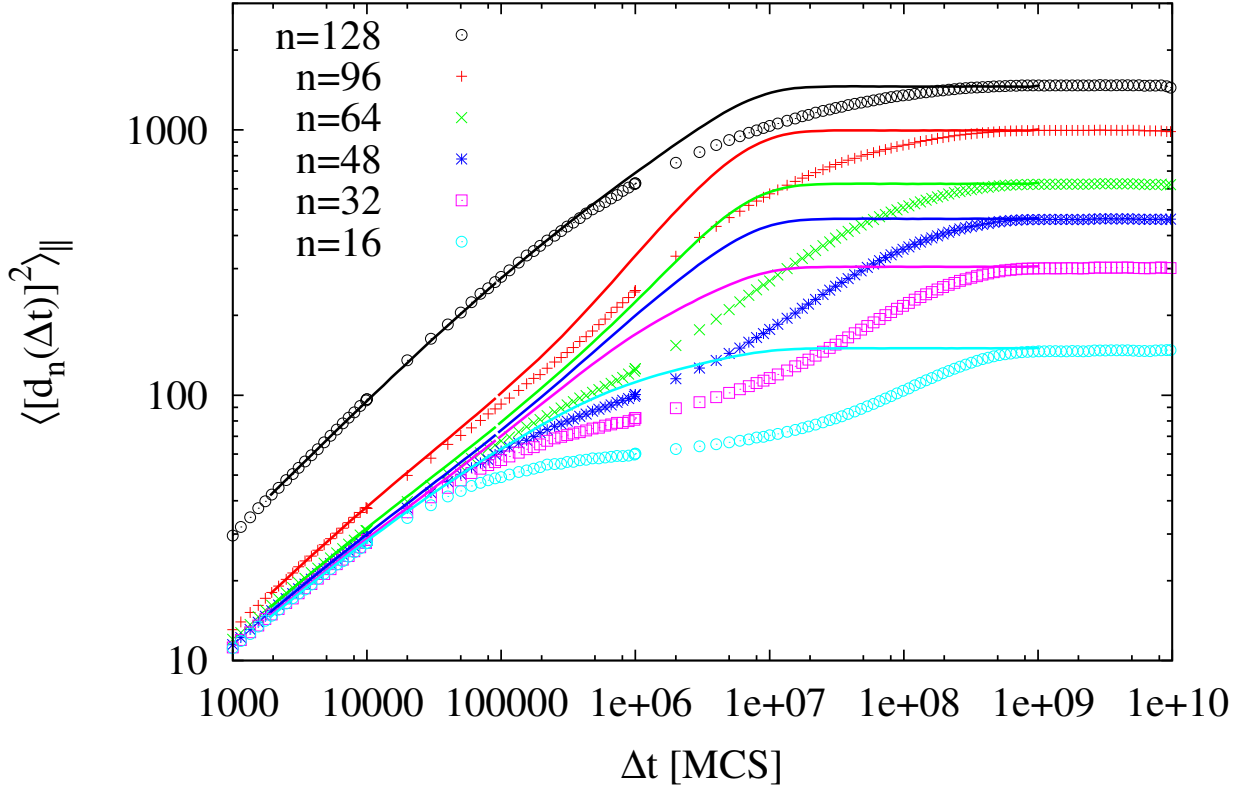}

\caption{\label{fig:Monomer-mean-square-1}Monomer mean square displacements
parallel to the grafting plane, $\left\langle \left[d_{n}(\Delta t)\right]^{2}\right\rangle _{\parallel}$,
for different monomers $n$ of sample \#7.}
\end{figure}

In Figures \ref{fig:Monomer-mean-square} and \ref{fig:Monomer-mean-square-1},
we show an example for the monomer mean square displacements as function
of $\Delta t$ for perpendicular and parallel directions separately.
For comparison with the numerical solution of a grafted Rouse chain
of Refs. \cite{Reith,Hoffmann-1}, the theoretical lines were shifted
for each $n$ individually along the time axis to match the displacements
near $\left\langle \left[d_{n}(\Delta t)\right]^{2}\right\rangle _{\perp}\approx\xi^{2}$
and along the displacement axis to match the pleateau at large $\Delta t$.
These shifts compensate differences in the average friction coefficients
and the different extensions of subsections of the chains as function
of $n$. 

In case of non-entangled brushes, we expect a good agreement between
these theoretical lines and the data in \emph{parallel} directions,
while the agreement should be poor in \emph{perpendicular} directions,
since $\left\langle \left[d_{n}(\Delta t)\right]^{2}\right\rangle _{\perp}\propto t^{2\nu/(1+2\nu)}\propto t^{2/3}$
for the self-similar relaxation of a swollen Rouse chain of size $R\approx bN^{\text{\ensuremath{\nu}}}$
with $\nu=1$. For entangled brushes we expect qualitatively the opposite
trend: poor agreement in parallel directions, since the monomers are
confined inside the tube, while the data in perpendicular direction
should follow Rouse dynamics up to the Rouse time of the chain, if
the entanglements damp out the enlarged fluctuations in the perpendicular
direction. The Rouse regime could be followed by a logarithmic regime,
if the exponentially rare retractions of the chains contribute significantly
to the mean square displacements. 

The data of both Figures are a strong hint towards entangled dynamics.
Since we observe no extended fluctuations in perpendicular directions,
there are either only few entanglements per chain such that tube length
fluctuations essentially dominate the perpendicular fluctuations or
the exponentially rare retractions do not significantly contribute
in perpendicular direction. Most characteristic for the data in parallel
directions is the extended cross-over from initial mean square displacements
(approximately $\propto t^{1/2}$) to a delayed equilibrium of mean
square displacements and the two apparent relaxation steps of $\left\langle \left[d_{n}(\Delta t)\right]^{2}\right\rangle _{\parallel}$
at small $n$.

\begin{figure}
\includegraphics[width=1\columnwidth]{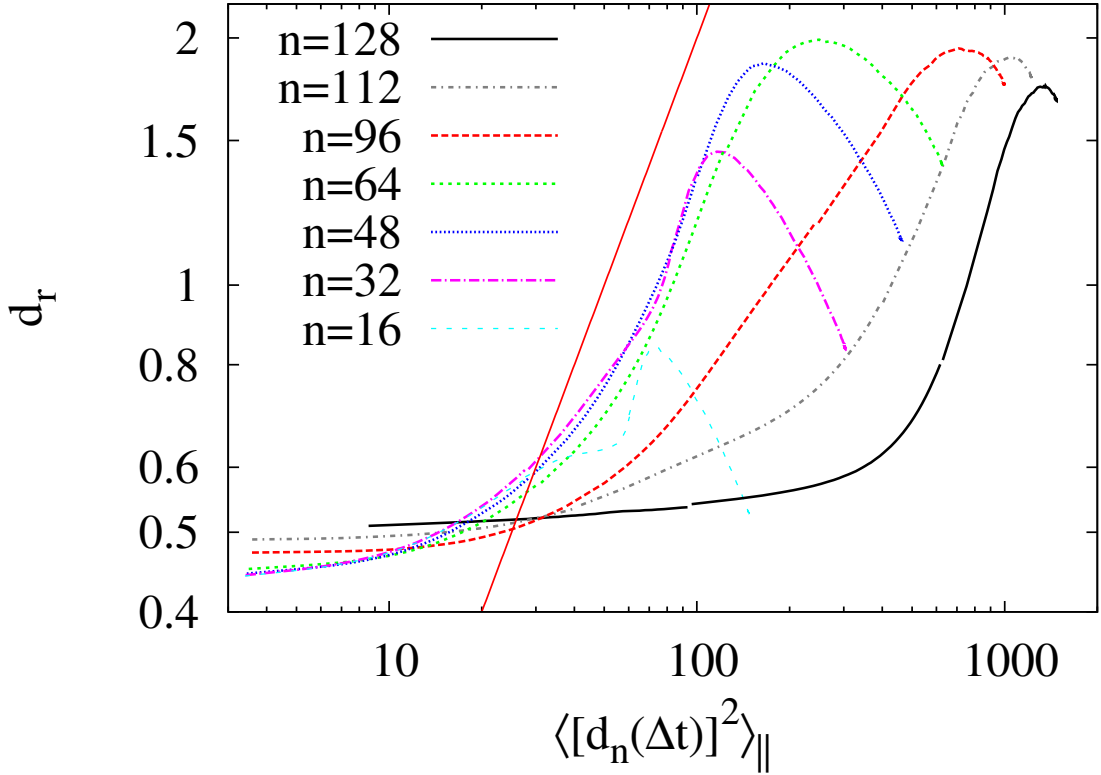}

\caption{\label{fig:Ratio--of}Ratio $d_{r}$ of the monomer mean square displacements
in perpendicular and parallel directions (see Figures \ref{fig:Monomer-mean-square}
and \ref{fig:Monomer-mean-square-1}) as function of the displacement
in parallel direction. The straight line indicates $d_{r}\propto\left\langle \left[d_{n}(\Delta t)\right]^{2}\right\rangle _{\parallel}$.}
\end{figure}

For a better qualitative understanding of the data, we plot the ratio
\begin{equation}
d_{r}\equiv\left\langle \left[d_{n}(\Delta t)\right]^{2}\right\rangle _{\perp}/\left\langle \left[d_{n}(\Delta t)\right]^{2}\right\rangle _{\parallel}\label{eq:dn-1}
\end{equation}
as function of $\left\langle \left[d_{n}(\Delta t)\right]^{2}\right\rangle _{\parallel}$
in Figure \ref{fig:Ratio--of}. This compensates the effects of a
different monomer mobility and allows for a more striking distinction
\cite{footnote} between the models discussed in the introduction.

All models \cite{Witten,Klushin,Johner,OConnor,OConnor2,Rubinstein-1}
start with an isotropic local motion of the monomers on a length scale
below blob size. Thus, we expect constant displacement ratios of 
\begin{equation}
d_{r}=1/2\label{eq:dn}
\end{equation}
for small $\left\langle \left[d_{n}(\Delta t)\right]^{2}\right\rangle _{\parallel}\lesssim\xi^{2}$
and $\tau<\tau_{\xi}$. Monomer diffusion becomes anisotropic for
an anisotropically swollen Rouse chain beyond the blob size with $\left\langle \left[d_{n}(\Delta t)\right]^{2}\right\rangle _{\perp}\propto t^{2/3}$,
leading to 
\begin{equation}
d_{r}\approx\frac{1}{2}\left\{ \left\langle \left[d_{n}(\Delta t)\right]^{2}\right\rangle _{\parallel}/\xi^{2}\right\} ^{1/3}\label{eq:dn2}
\end{equation}
for $t<\tau_{\parallel,d}(n)$ or 

\begin{equation}
\left\langle \left[d_{n}(\Delta t)\right]^{2}\right\rangle _{\parallel}\lesssim\frac{\xi^{2}n}{g}.\label{eq:dnpar}
\end{equation}
For $\tau_{\parallel,d}(n)<t<\tau_{\perp,d}(n)$, only the mean square
displacements in perpendicular directions grow up to

\begin{equation}
\left\langle \left[d_{n}(\Delta t)\right]^{2}\right\rangle _{\perp}\approx\frac{\xi^{2}n^{2}}{g^{2}}.\label{eq:dnper}
\end{equation}
Thus, we obtain three regimes for the non-entangled dynamics of polymers
swollen in the stretching field of a brush: $d_{r}=const.$ followed
by a power law with power 1/3 and an individual, sharp vertical increase
of $d_{r}$ for each $n$ at the corresponding maximum mean square
displacement in parallel directions.

In case of arm retraction \cite{Witten}, the initial regime, equation
(\ref{eq:dn}) is obtained up to $\left\langle \left[d_{n}(\Delta t)\right]^{2}\right\rangle _{\parallel}\lesssim\xi^{2}$.
For larger displacements, the motions of inner monomers are confined
within an ``over-stretched'' tube of diameter $\xi$. This tube
is similarly stretched as the chain conformations in the brush, and
thus, performs a random walk in parallel directions, while it can
be considered as a fully stretched array of entangled sections in
perpendicular direction. This geometrical constraint leads to 
\begin{equation}
d_{r}\propto\left\langle \left[d_{n}(\Delta t)\right]^{2}\right\rangle _{\parallel}\label{eq:dr}
\end{equation}
for $\xi^{2}\lesssim\left\langle \left[d_{n}(\Delta t)\right]^{2}\right\rangle _{\perp}\lesssim\xi^{2}n^{2}/(3g^{2})$.
Full exploration in parallel directions requires a sufficient number
of full retractions, while these exponentially rare retractions contribute
only little to the mean square displacements in perpendicular direction.
Thus, the limiting $\left\langle \left[d_{n}(\Delta t)\right]^{2}\right\rangle _{\parallel}$
at very large $\Delta t$ should appear delayed as compared to the
mean square displacements in perpendicular direction near to the limiting
$\left\langle \left[d_{n}(\Delta t)\right]^{2}\right\rangle _{\perp}$.
Therefore, while $\left\langle \left[d_{n}(\Delta t)\right]^{2}\right\rangle _{\perp}$
is roughly constant, $\left\langle \left[d_{n}(\Delta t)\right]^{2}\right\rangle _{\parallel}$
remains growing, which implies a broad cross-over from $d_{r}\propto\left\langle \left[d_{n}(\Delta t)\right]^{2}\right\rangle _{\parallel}$
to a dependence 
\begin{equation}
d_{r}\propto\left\langle \left[d_{n}(\Delta t)\right]^{2}\right\rangle _{\parallel}^{-1}\label{eq:dr2}
\end{equation}
near maximum mean square displacements in parallel directions.

We show the ratio $d_{r}$ for the mean square displacements of selected
monomers of sample \#7 in Figure \ref{fig:Ratio--of}. All data show
an initially constant $d_{r}\approx1/2$ as expected for isotropic
monomer motion. For essentially all inner monomers (small $n$), this
regime is followed by a regime with $d_{r}\propto\left\langle \left[d_{n}(\Delta t)\right]^{2}\right\rangle _{\parallel}$
until the data turn into $d_{r}\propto\left\langle \left[d_{n}(\Delta t)\right]^{2}\right\rangle _{\parallel}^{-1}$
near the largest displacements. Thus, Figure \ref{fig:Ratio--of}
demonstrates that the chains of this brush are clearly entangled.

The outermost monomers $n\approx128$ are partially in agreement with
the qualitative behavior predicted for an anisotropically swollen
Rouse chain. The large correlation blob size near the free end monomer
probably prevents the observation of the 1/3 power law predicted above.
A small downturn region is visible near the end, which indicates a
marginal impact of entanglements. For $n<128$, a gradual transition
to the behavior of the entangled inner monomers is observed, see $n=96,$
112 for examples.

For a quantitative test of our discussion above concerning tension
blobs that are smaller than tube diameter, we extrapolate the data
for small $n$ with $d_{r}\propto\left\langle \left[d_{n}(\Delta t)\right]^{2}\right\rangle _{\parallel}$
towards a level of $d_{r}=1/2$. We find an intersection point at
a mean square displacement of $37\pm3$ square lattice units for all
$N/4<n<N/2$, which is identical within the error bars with the average
square distance between the grafting points, $\xi^{2}$. Thus, the
entanglements are pulled taught by the large tension along the chains
and we can consider each polymer strand roughly as an independent
obstacle where other chains have to wind around. With $2b^{2}/3\approx4.7$
for our simulation model in parallel directions, we estimate for the
average number of monomers between two consecutive contacts (or the
mean distance between slip-links - not to be confused with tube diameter
\cite{Lang}), $N_{p}$, that $N_{p}\approx8$ for $\sigma\approx1/9$
in sample \#7. Note that the above extrapolation of $d_{r}$ is the
dynamic equivalent of determining blob size from polymer conformations.
The good agreement of the intersection point with $\xi^{2}$ allows
us to estimate the number of monomers per ``dynamic blob'' by setting
$g=(\xi/b)^{1/\nu}$ without additional numerical coefficients. This
relation is used below for a quantitative analysis of the relaxation
times.

As bottom line of the above section we conclude that the relaxation
in perpendicular and parallel directions must be coupled by confinement
in a tube; at least we have no other explanation for the behavior
of $d_{r}$ in Figure \ref{fig:Ratio--of}. The slow approach of the
limiting mean square displacements in Figure \ref{fig:Monomer-mean-square}
and \ref{fig:Monomer-mean-square-1} makes it rather difficult to
conclude on terminal relaxation times using mean square displacement
data. In the following section, we discuss how to analyze the terminal
relaxation time of a polymer brush using auto-correlation functions.

\subsection{\label{sub:Auto-correlation-functions}Position vector auto-correlation
function and terminal relaxation times}

\begin{figure}
\includegraphics[width=1\columnwidth]{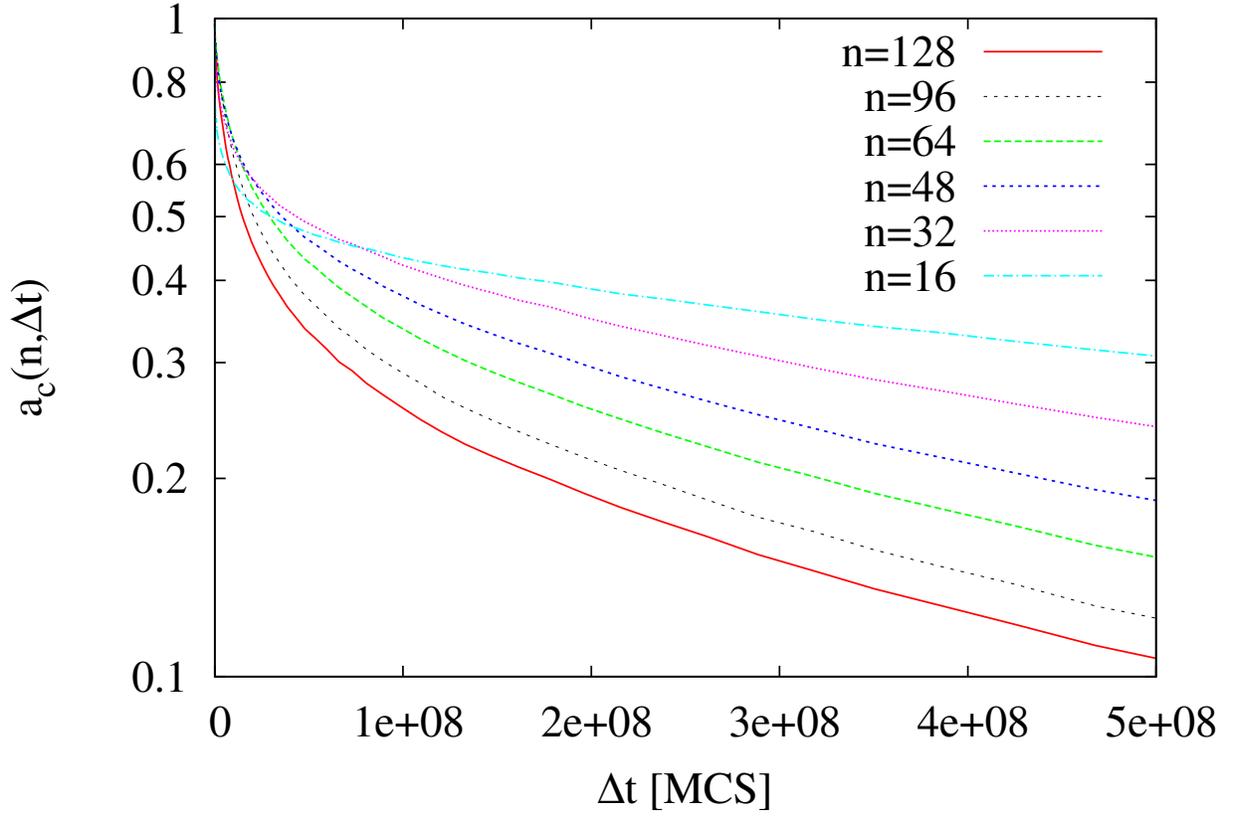}

\caption{\label{fig:Isotropic-attachment-vector}Isotropic position vector
auto-correlation function $a_{c}(n,\Delta t)$ for sample \#6.}
\end{figure}

The isotropic position vector auto-correlation function $a_{c}(n,\Delta t)$
given by equation (\ref{eq:ac}) is plotted in Figure \ref{fig:Isotropic-attachment-vector}
in the standard way (semi-log plot with linear time axis) for analyzing
linear polymers. As expected from the previous section, there is no
dominating single exponential decay as a function of time that can
be used to identify a terminal relaxation. Note that such an approach
also fails for the outermost monomers with largest $n$. Taking a
decay to a fixed level of, for instance, $1/e$ as criterion for determining
the relaxation time (using EAT, CAT, RAT, or DS, see Table \ref{tab:Scaling-of-terminal})
misses the increasingly slowed down relaxation at longer times, see
Figure \ref{fig:Isotropic-attachment-vector}.

\begin{figure}
\includegraphics[width=1\columnwidth]{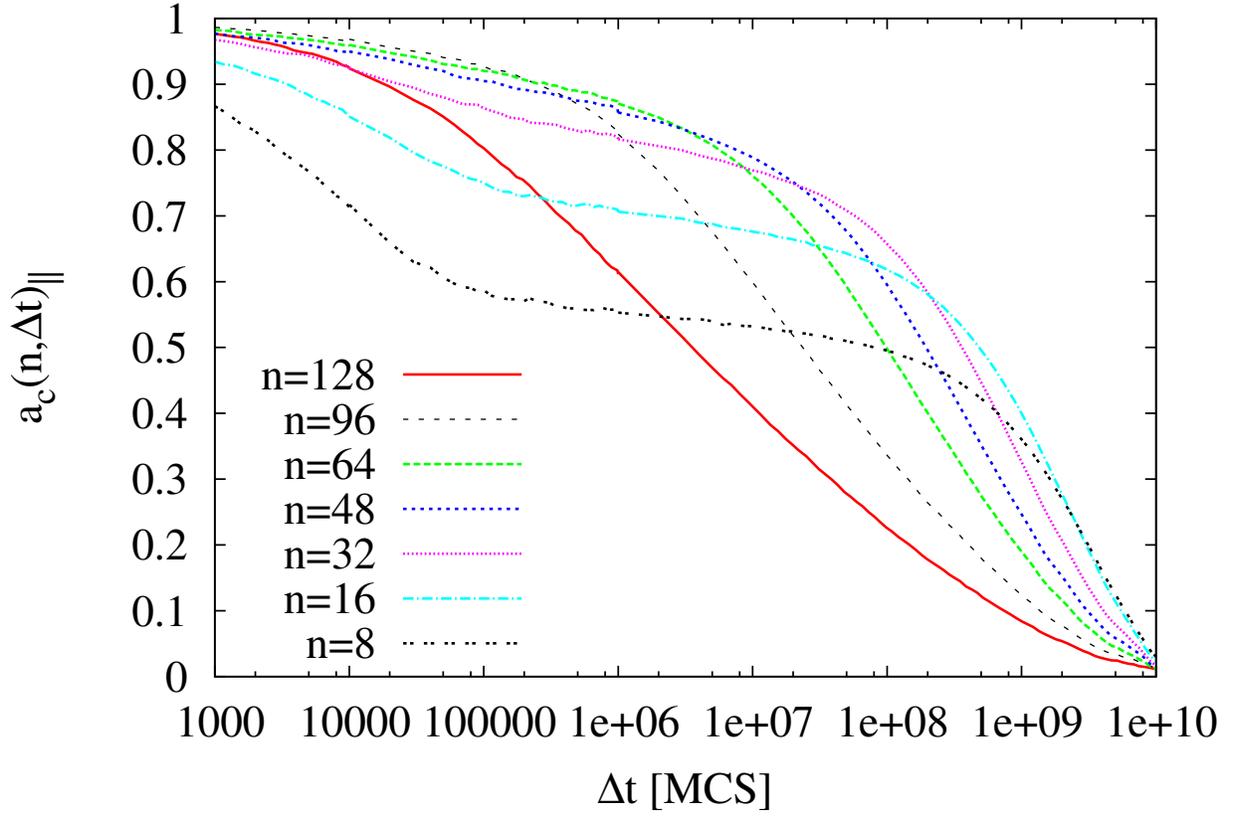}

\caption{\label{fig:Attachment-vector-auto-correlati-1}Position vector auto-correlation
function $a_{c}(n,\Delta t)_{\parallel}$ in parallel directions for
sample \#6.}
\end{figure}

\begin{figure}
\includegraphics[width=1\columnwidth]{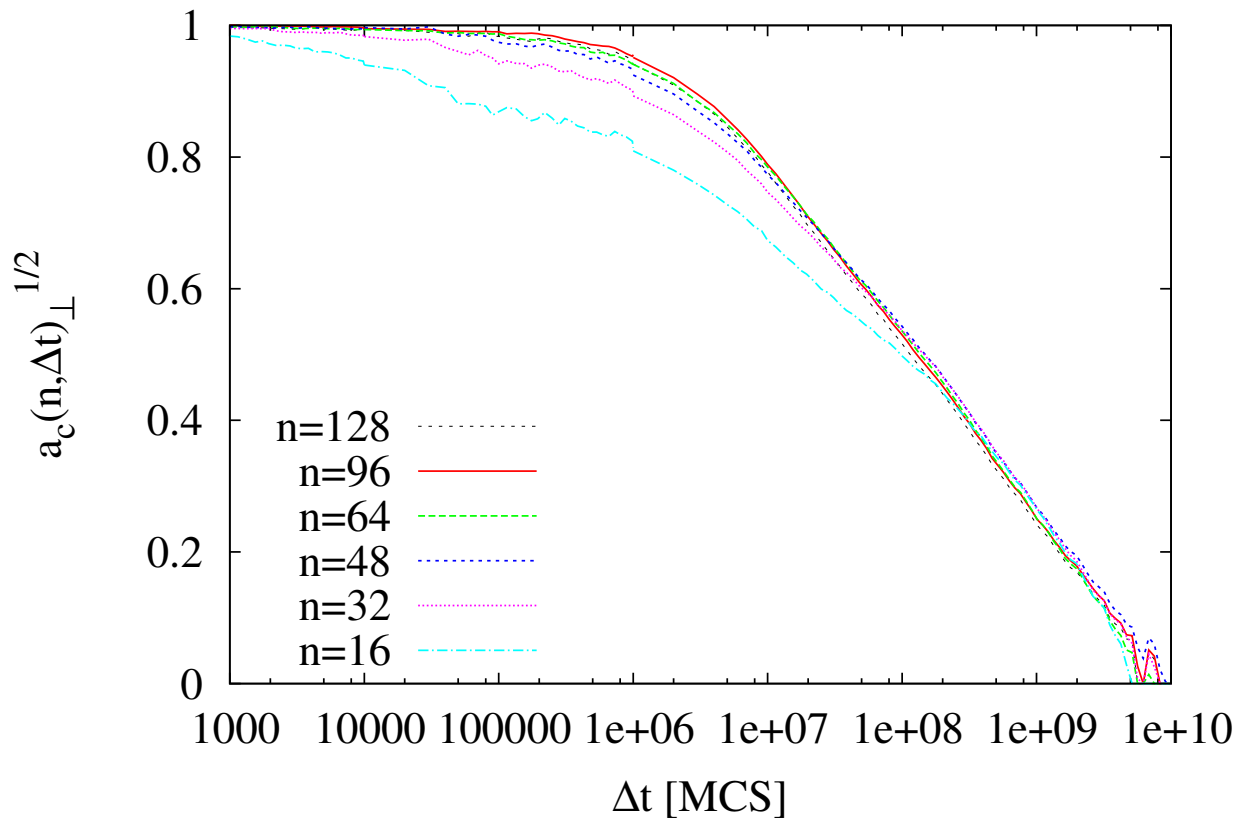}

\caption{\label{fig:Attachment-vector-auto-correlati-2}The square root of
the Position vector auto-correlation function $a_{c}(n,\Delta t)_{\perp}$
in perpendicular directions for sample \#6 indicates the portion of
the chain that is not yet relaxed at time $\Delta t$.}
\end{figure}

Let us now discuss the directional components of the auto-correlation
function in perpendicular and parallel directions separately (Figures
\ref{fig:Attachment-vector-auto-correlati-1} and \ref{fig:Attachment-vector-auto-correlati-2})
using a logarithmic time axis as suitable for understanding activated
relaxation, see equation (\ref{eq:s(t)}). Again, we observe clear
qualitative differences between parallel and perpendicular directions.
The data of the inner monomers (small $n$) of the parallel directions
in Figure \ref{fig:Attachment-vector-auto-correlati-1} follow qualitatively
the expectations of arm retraction models: after an initial quick
Rouse-like decay, there is a logarithmic plateau that is stable for
several decades in time until the innermost tube sections relax. The
data at largest times show that terminal relaxation occurs hierarchically
from the free end to the innermost monomers. Note that Figure \ref{fig:Attachment-vector-auto-correlati-1}
fully reproduces the observation of Ref. \cite{Reith}, where the
largest relaxation times could be found near the middle of the chains:
since we show the data of a brush with somewhat larger $N/g$ as discussed
by Reith et al. \cite{Reith}, such a behavior is found for a larger
level of approximately $0.7$ for the correlation function (a smaller
portion of Rouse-like relaxation for the more entangled chains of
the present study shifts the logarithmic plateaus to higher levels).
Furthermore, Figure \ref{fig:Attachment-vector-auto-correlati-1}
shows that even the less suitable, analysis of a decay to $1/e$ can
reveal a hierarchical relaxation for the parallel components, if the
chains are sufficiently entangled. Nevertheless, such an approach
leads always to a systematic underestimation of the terminal relaxation
time.

The perpendicular component of the position vector auto-correlation
function, which is the dominating contribution for long chains in
the brush regime, is shown in Figure \ref{fig:Attachment-vector-auto-correlati-2}.
Note that we plot the square root of the correlation function, which
we use below for the determination of the terminal relaxation time.
The main difference to the parallel directions is that final relaxation
is reached \emph{non-hierarchically}. But in contrast to non-entangled
dynamics, relaxation occurs with a \emph{logarithmic time dependence},
showing key signatures of self-similar dynamics \emph{and} retractive
motion all the way up to the terminal relaxation. In principle, such
a relaxation is only possible, if there is a mechanism that couples
the hierarchical relaxation as a function of $n$ in perpendicular
direction, while leaving the parallel directions mainly untouched.
This is particularly surprising, since the mean square displacement
data show that we have to consider the relaxation of chains confined
in a tube like environment, and thus, the relaxation of \emph{tube}
\emph{sections} in parallel and perpendicular directions must occur
simultaneously. Below, we show by a stepwise generalization of a simple
model case that exactly such a hybrid relaxation must follow from
what is known about arm retraction dynamics and chain conformations
in a brush.

Let us first consider an ideal polymer chain that is stretched by
a self-consistent potential that models the interactions with other
chains in a brush. This chain is additionally confined by a perfectly
straight solid tube of uniform diameter with orientation perpendicular
to the grafting plane. In this case, relaxation in parallel direction
is finished at the exploration time of the tube diameter, while the
perpendicular direction relaxes at the Rouse time of the non-entangled
chain in the brush potential.

Next, let us consider that the straight tube is tilted by an angle
$\theta$ away from the perpendicular direction and let us re-normalize
the brush potential such that the time average monomer positions in
the coordinate system of the tilted tube do not change. In this case,
we obtain the same relaxation behavior as before up to small geometrical
coefficients reflecting the tilted tube geometry. We conclude that
a constant average tilt of the tube sections does not impose a qualitatively
different relaxation behavior of the chain.

The same result is obtained, if we split the tube into $s_{max}$
straight sections of identical length, which are tilted by the same
small angle $\theta$ away from the perpendicular axis but rotated
indiviually around this axis. In this case, the parallel components
of these tube sections resemble a random walk, while the perpendicular
component is a directed walk away from the grafting point. Note further
that relaxation is also not modified, if we allow for a fluctuation
of the length of these tube sections while keeping $\theta$ fixed.
The reason for this observation is that in all of the above cases
we find the very same value of the brush potential at the very same
coordinate along the axis of the confining tube.

In the next step of generalization, we randomly distort the individual
tilt angles of the tube sections such that the average perpendicular
coordinate of tube section $s$ in a large ensemble of randomly distorted
tubes is not modified. Thus, there is still the same value of the
brush potential at the very same ensemble average coordinate of the
confining tubes. However, this relation no longer holds when analyzing
individual tubes. In consequence, the time average force acting on
the monomers (and thus, the time average monomer positions in perpendicular
direction) becomes a function of the particular tube conformation.

Let us now replace each kink of these individual tubes with a slip
link for confining the fluctuations of the chain. Relaxation of the
polymer requires in this case that the free end finds its way back
through the particular set of slip links. This modification introduces
the exponentially growing relaxation times for deep retractions discussed
at the introduction, since now the extra chain conformations outside
of the slip-link lead to an entropic penalty for retraction. Note
that the brush potential with respect to a particular conformation
of such a ``tube'' is sampled on the much shorter Rouse time of
this ``stretched'' chain such that modifications of sections $s\lesssim s_{max}-\left(N/N_{e}\right)^{1/2}$
are fully communicated all along the chain contour prior to the following
relaxation steps.

To proceed, we have to make our simple model more self-consistent.
To this end, we consider that deviations from the average monomer
positions are driven by thermal agitation. Since the confining tube
(i.e. the slip link positions) must follow the same statistics as
the polymer conformations, an average orientation fluctuation of section
$s$ stores an amount of $\approx kT$ of energy. This energy becomes
available for sampling extra conformations (different orientations
of tube section $i\ge s$) after the corresponding entanglement has
been released. Since these modifications are fully communicated in
the retraction regime, each relaxing tube sections leads roughly to
the same drift of the average monomer positions in perpendicular direction
as a result of the modified time average force acting on the monomers
independent of the particular position of $s$.

Finally, we have to take into account that the brush potential causes
extra large fluctuations of monomer positions $\propto n\propto s$
around the average positions in perpendicular direction. Thus, the
effect on the drift of average monomer positions due to orientation
fluctuations is proportional to the weight fraction $s$ of the not
yet relaxed tube sections, which enter squared in $a_{c}(n,\Delta t)_{\perp}$.
Thus, $a_{c}(n,\Delta t)_{\perp}^{1/2}$ must be $\propto s(t)$ at
large times. According to equation (\ref{eq:s(t)}), $s(t)$ decays
linearly when plotted on logarithmic time axis. This is what we observe
in Figure \ref{fig:Attachment-vector-auto-correlati-2}.

We must explain also, why the drift of the average monomer positions
in perpendicular direction is not dominating the relaxation in parallel
directions, since relaxation of the different directions is coupled
by the curvilinear motion of the chain along the confining tube. Notice
that the mean square displacements in perpendicular direction take
maximum values of approximately $\approx(cn/N)^{2}$ with a small
coefficient, for instance, $c\approx0.23H$ for a parabolic brush
\cite{Klushin}. Ignoring the effect of tube length fluctuations,
the monomers, thus, move less than 1/4 along the tube in parallel
directions as a result of the drift in perpendicular direction (since
this drift is competing with tube length fluctuations, the actual
contribution is smaller). The remaining 3/4 of the tube in parallel
directions relax by arm retraction. Thus, the coupling in perpendicular
direction enters only as a correction that causes a weak decay of
the correlation plateau of innermost monomers in Figure \ref{fig:Attachment-vector-auto-correlati-1}
together with additional modes of constraint release \cite{Viovy},
dynamic dilution \cite{Ball}, or tube dilation \cite{Marucci} due
to the relaxation of surrounding chains. 

Note that we neglected above a possible distribution in the number
of entanglements per chain as discussed previously for the stress-relaxation
of block co-polymers \cite{Rubinstein-1}. This is motivated from
the fact even for that small $N/N_{e}$ of the present study we do
not observe a dramatic decay of the correlation plateau at intermediate
times (see above). Furthermore, this distribution is expected to be
of Poisson type and thus, vanishes for $N/N_{e}\gg1$ with the same
dependence as tube length fluctuations.

Therefore, in order to determine the relaxation time of an entangled
brush, we simply fit all data with $a_{c}(n,\Delta t)_{\perp}^{1/2}<1/4$
using functions of the form $\alpha_{n}\log(\tau_{n}/\Delta t)$.
Here, $\alpha_{n}$ is the slope of the decay and $\tau_{n}$ the
terminal relaxation time of $a_{c}(n,\Delta t$). The relaxation time
$\tau$ of each brush is defined as the average $\tau_{n}$ obtained
by this procedure and is reported in Table \ref{tab:Simulation-parameters-for}.
Note that for all samples, except of the least entangled samples \#1
and \#5, we observe a rather unique logarithmic terminal decay independent
of $n$ at large times similar to the one shown in Figure \ref{fig:Attachment-vector-auto-correlati-2}.

A plot of $a_{c}(n,\Delta t)_{\parallel}^{1/2}(n/N)^{1/2}$ (taking
into account that only a fraction $n/N$ instead of $N$ monomers
contribute to the not yet relaxed tube sections) shows that the terminal
relaxation in parallel directions occurs simultaneously (see also
Figure \ref{fig:Attachment-vector-auto-correlati-1}) for all $n$
somewhat delayed at 2-3 $\tau$ , indicating that about 2-3 dives
of the full chain are necessary to explore space in parallel directions.
Thus, the terminal relaxation times coincide up to a coefficient of
order unity, as expected for chains confined in a tube and in contrast
to non-entangled dynamics, see equations (\ref{eq:tauperp}) and (\ref{eq:tauparallel})
or chains in an array of rods. This observation also rules out that
relaxation is driven by a disentanglement in parallel directions (cf.
Introduction) for brushes with parameters as in the present study.

\begin{figure}
\includegraphics[width=1\columnwidth]{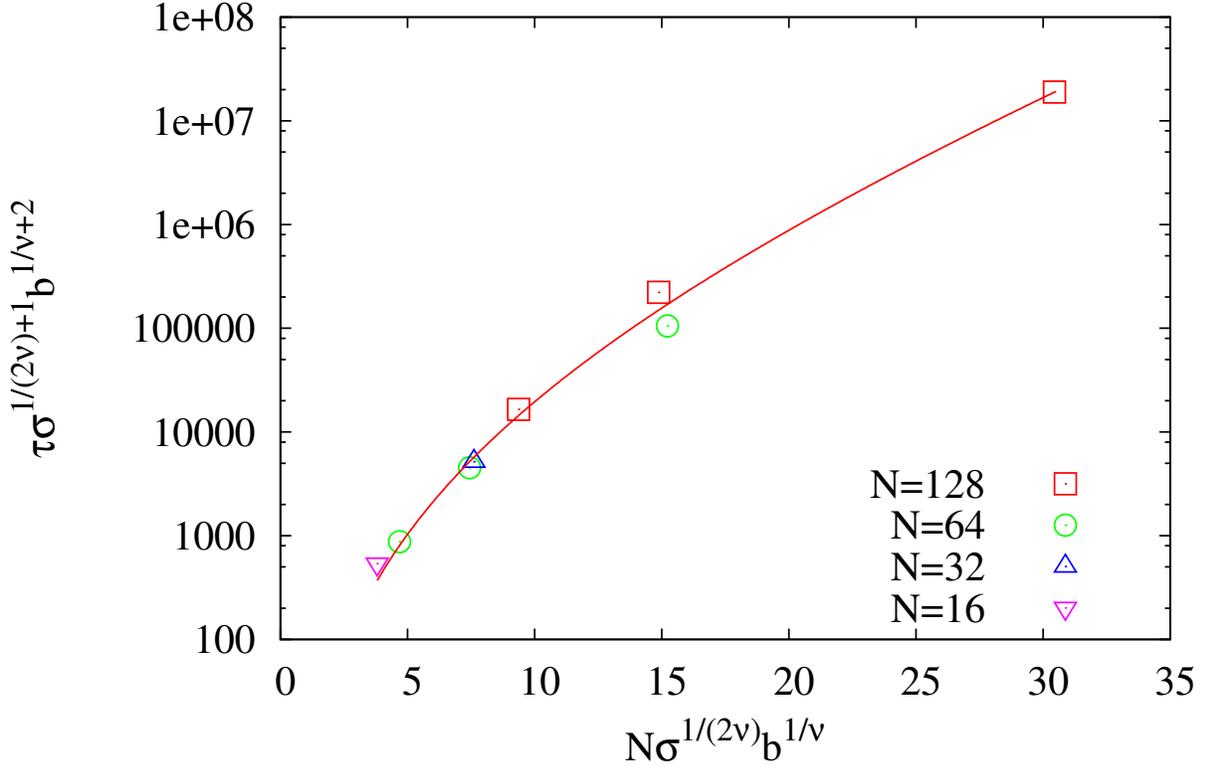}

\caption{\label{fig:terminalrelaxation}Dynamic scaling of the terminal relaxation
times as a function of the number of blobs $N/g\approx N\sigma^{1/(2\nu)}b^{1/\nu}$
and as multiple of the Rouse time $\tau_{\xi}$ of a blob, $\tau/\tau_{\xi}\approx\tau\sigma^{1/(2\nu)+1}b^{1/\nu+2}$.
The line is a fit to equation (\ref{eq:relax}) as described in the
text.}
\end{figure}

The data for terminal relaxation times of Table \ref{tab:Simulation-parameters-for}
are plotted in Figure \ref{fig:terminalrelaxation}. We observe a
reasonable overlap, if we rescale $N$ by the number of monomers per
dynamic blob, $N/g=N\sigma^{1/(2\nu)}b^{1/\nu}$, as determined from
monomer mean square displacements and relaxation times using the relation
for the Rouse time of a blob, equation (\ref{eq:taux}). The data
are fit with a function of form $\tau\propto(N/g)^{3}\exp(\beta N/N_{e})$
as expected for dense, entangled polymer brushes (cf. equation (\ref{eq:relax})).
Since the data are plotted as function of $N/g$, we can estimate
from the coefficient $\beta\approx0.17\pm0.02$ that $N_{e}/g\approx5.9\pm0.8$.
This value is about half of what has been conjectured for flexible
polymers in melts \cite{Lin,Kavassalis}. The mean square displacement
data in the previous section also indicated that the degree of polymerization
between consecutive ``slip links'', $N_{p}$, is $\approx g$. Thus,
we obtain $N_{e}/N_{p}\approx6$, which is clearly larger as what
is typically considered in melts or networks with estimates ranging
from $2$ in Ref. \cite{Everaers} over $7/4$ for the slip tube model
\cite{Rubinstein} to $5/4$ in Ref. \cite{Larson}, for instance.
Note further that with $N/g\approx30$ for sample \#6, we obtain $N/N_{e}\approx5\pm1$.
Finally, we want to emphasize that a smaller $N_{e}/g\approx6$ for
a larger overlap number $P\propto(N/g)$ in brushes as compared to
a melt of blobs with $P\propto(N/g)^{1/2}$ is not surprising, if
entanglement can be considered as a consequence of polymer overlap.
Some part of the qualitative changes (smaller $N_{e}/g$ and larger
$N_{e}/N_{p}$) also might be explained by the over-stretched conformations
inside the tube, which causes fewer contacts of higher energy with
only some of the overlapping chains.

\subsection{\label{sub:Auto-correlation-functions-1}Hierarchical relaxation
as analyzed via bond auto-correlation functions}

For a local analysis of hierarchical relaxation we use the bond auto-correlation
functions $b_{c}(n,\Delta t)$ defined in equation (\ref{eq:bc}).
Below, we focus only on the parallel directions in order to minimize
the perturbing effect of the coupling in perpendicular direction.
For times below the blob relaxation time, equation (\ref{eq:taux}),
each blob relaxes as part of a swollen Rouse chain to a correlation
level of $\xi^{2}/(b^{2}g^{2})\approx g^{2\nu-2}$. Since this occurs
on the relaxation time $\propto(t/\tau_{0})^{1+2\nu}$ of a swollen
Rouse chain, we expect the bond correlation function $b_{c}(n,\Delta t)$
to decay as 
\begin{equation}
b_{c}(n,\Delta t)\propto\left(\frac{\Delta t}{\tau_{0}}\right)^{-(2-2\nu)/(1+2\nu)}\propto\Delta t^{-0.38}.\label{eq:bc-1}
\end{equation}

Since the inner chain monomers reside in ``over-stretched'' tubes,
we expect beyond $\tau_{\xi}$ and before the Rouse time 
\begin{equation}
\tau_{n}\approx\tau_{\xi}\left(\frac{n}{g}\right)^{2}\label{eq:taun-1}
\end{equation}
of a section of $n$ monomers, $\Delta t<\tau_{n}$, a regime with
\begin{equation}
b_{c}(n,\Delta t)_{\parallel}\propto\left(\frac{\Delta t}{\tau_{e}}\right)^{-1/4},\label{eq:bcpar-1}
\end{equation}
similar to entangled chain sections in melt. This regime is followed
by an almost constant entanglement plateau (as a function of time)
below the terminal relaxation time of a bond, $\tau_{b}$, that might
be affected by tube dilation or constraint release \cite{Marucci,Ball,Viovy}
or the drift along the tube as function of the relaxing tube sections.
The terminal time can be estimated by the retraction time of the chain
end along a path of length proportional to the distance to the free
end. Assuming a retraction path length $\propto(N-n)$ at constant
blob size in perpendicular direction, for simplicity, we obtain 
\begin{equation}
\tau_{b}(n)\approx N(N-n)^{2}/g^{3}\exp\left(\frac{(N-n)^{2}}{NN_{e}}\right)\label{eq:taub}
\end{equation}
using the estimate of Witten et al. \cite{Witten} and replacing $Ng$
by $NN_{e}$ in the exponential argument, as above. Note that this
entanglement plateau is only possible if $\tau_{n}<\tau_{b}(n)$.

\begin{figure}
\includegraphics[width=1\columnwidth]{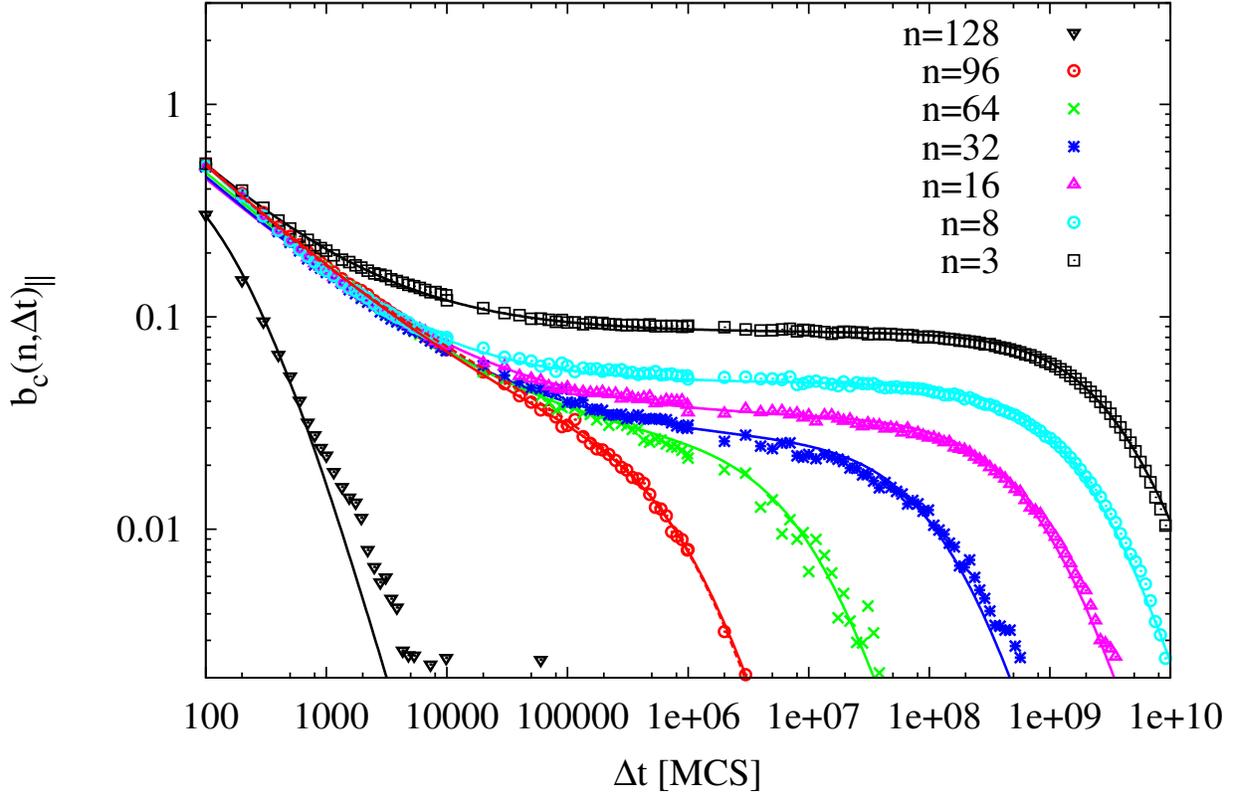}

\caption{\label{fig:Bond-auto-correlation-function}Bond auto-correlation function
in parallel directions for sample \#6. The continuous lines are fits
to the data as discussed in the text. }
\end{figure}

Figure \ref{fig:Bond-auto-correlation-function} shows a well pronounced
plateau for the innermost monomers (small $n$). However, the small
values of $n/N_{e}$, $N_{e}/g$, and $g$ that are possible for $n<N\le128$
(if there is a significant number of entangled sections $N/N_{e}\gg1$
and ignoring the effect of tension blobs smaller than tube diameter)
do not allow for an unambiguous observation of the early power law
regimes. For instance, an apparent regime with $b_{c}(n,\Delta t)\propto\Delta t^{-1/4}$
over about one decade in time may also be caused by a power law decay
followed by a short plateau. To demonstrate this point we include
two different fits for the data of $n=96$ of the most entangled sample
\#6. The first fit function reads 
\begin{equation}
f_{1}(t)=\left(c_{1}t^{-0.38}+c_{2}t^{-1/4}\right)\left(\tau_{b}/(t+\tau_{b})\right)^{2},\label{eq:fit1}
\end{equation}
while we used 
\begin{equation}
f_{2}(t)=\left(c_{1}t^{-x}+c_{2}\right)\left(\tau_{b}/(t+\tau_{b})\right)^{2},\label{eq:fit2}
\end{equation}
with a variable $1/4\le x\le0.38$ in second place. Both functions
lead to equally well fits of the data for $n$ roughly in the range
of $96\pm12$, see $n=96$ in Figure \ref{fig:Bond-auto-correlation-function}
as example. However, the terminal times $\tau_{b}$ obtained can differ
up to a factor of two. Notice that $f_{2}(t)$ fits clearly better
for $n<84$, while $f_{1}(t)$ was used for $n>108$ with an essentially
vanishing second power law for $n\rightarrow N$ and that we use a
square cut-off $\left(\tau_{b}/(t+\tau_{b})\right)^{2}$ to determine
$\tau_{b}$ for all monomers $n<N$ .

\begin{figure}
\includegraphics[width=1\columnwidth]{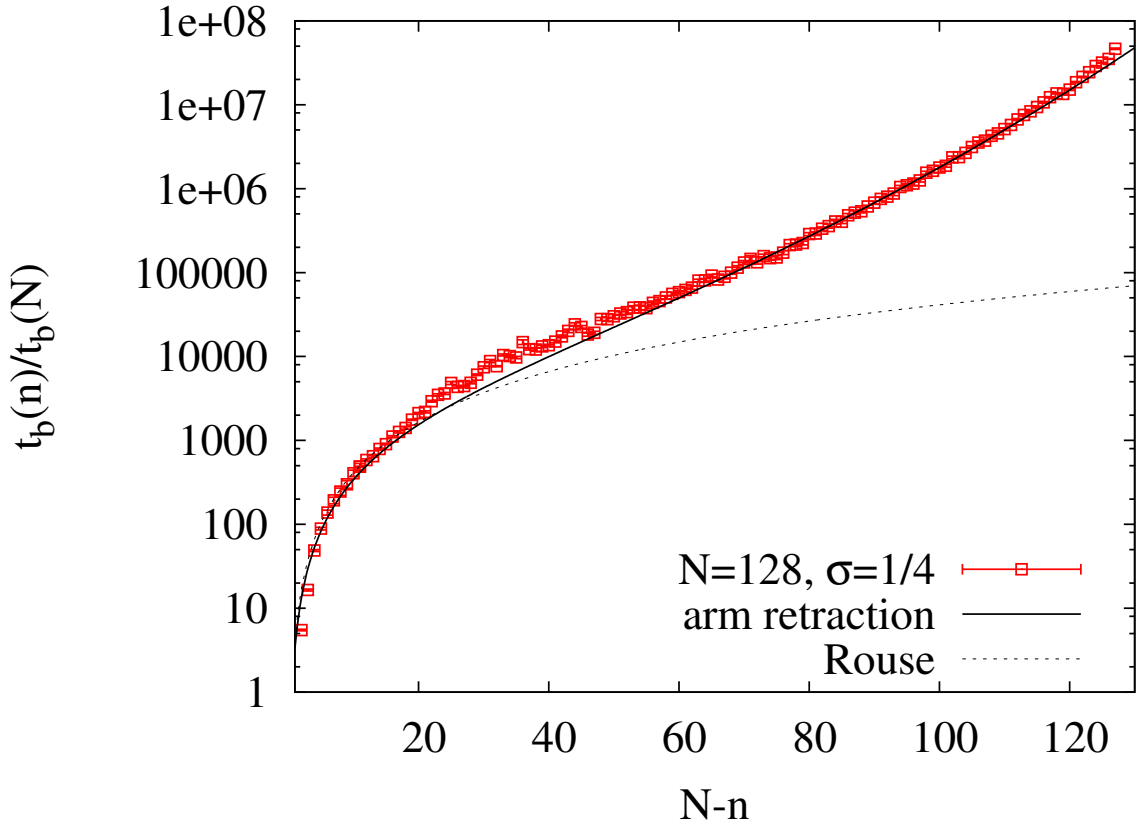}

\caption{\label{fig:Simple-fit-to}Terminal bond relaxation times $\tau_{b}(n)$
in perpendicular direction as estimated from fitting equations (\ref{eq:fit1})
and (\ref{eq:fit2}) to the data of sample \#6 with $N=128$ and $\sigma=1/4$,
see text for details. The continuous line (arm retraction) is a fit
to equation (\ref{eq:taub}), while the dashed line is a fit according
to $\left(N-n\right)^{2}$ for small $N-n$.}
 
\end{figure}

Figure \ref{fig:Simple-fit-to} shows the fit results for $\tau_{b}(n)$,
whereby we interpolate linearly between the results for $f_{1}(t)$
and $f_{2}(t)$ in the range $84<n<108$ in order to obtain a smooth
transition between the domains where both fit functions were used.
The difficulty to consistently fit the $\tau_{b}$ data in this transition
zone is also visible by the larger scatter of the data for $20\le N-n\le44$.
Nevertheless, the data at all $N-n$ fit well to equation (\ref{eq:taub}),
while only the small $N-n$ range near the end can be approximated
by relaxation times $\propto(N-n)^{2}$ as one expects from the Rouse
model in parallel direction. Note that we exclude the regions with
$N-n>120$ and $N-n<10$ from the first fit in order to reduce end
effects and that the second fit is applied for $N-n<20$. From the
fit to equation (\ref{eq:taub}) we obtain $N_{e}\approx20\pm1$,
which gives $N/N_{e}\approx6.4\pm0.3$ for sample \#6, which is in
fair agreement with our previous estimate of $N/N_{e}\approx5\pm1$
from Figure \ref{fig:terminalrelaxation}. Note that a similar agreement
with larger error bars for $N_{e}$ is found for the less entangled
samples of our study.

\section{Summary}

In this work, we analyzed in detail the relaxation of monodisperse
brushes. We demonstrated for the first time by computer simulations
that the relaxation of a polymer brush is governed by retraction dynamics
- even for the small number of correlation volumes per chain of the
present study. We have shown that this result can be missed with the
conventional analysis of the decay of the position vector auto-correlation
functions to a level of $1/e$. Such an analysis is suitable for self-similar
relaxation, but leads to significant errors in case of arm retraction,
since the logarithmic decay at very long times is not fully detected.

The terminal relaxation times of all brushes collapse on a master
curve, when time is rescaled by the relaxation time of a concentration
blob, while $N$ and the entanglement degree of polymerization, $N_{e}$,
are rescaled by the numbers of monomer per concentration blob, $g$.
For the relaxation times of all samples we find an exponential increase
$\tau\propto(N/g)^{3}\exp(N/N_{e})$ as a function of the degree of
polymerization, $N$. From the master plot we obtain $N_{e}/g\approx5.9\pm0.8$
and (for the most entangled sample of our study) $N/N_{e}\approx5\pm1$.
Thus, entanglement effects already dominate the long time dynamics
of a polymer brush much earlier as estimated previously \cite{Klushin}
or as compared to monodisperse melt or network data of the same simulation
model \cite{Lang}. This earlier onset of entanglement effects may
have several reasons. For instance, the overlap number grows $\propto N$
in a brush as compared to $\propto N^{1/2}$ in a melt, reptation
is impossible for grafted chains, and the hierarchical relaxation
in combination with the average chain orientation largely reduces
the impact of constraint release to surrounding polymers (overlapping
chain sections essentially relax at the same time scale). According
to our data, entanglements must be considered for $N/g\gtrsim10$,
which is roughly the same criterion as it is used for the application
of the strong stretching approximation for chain conformations \cite{Romeis}.

The bond auto-correlation functions in parallel directions show a
strictly hierarchical relaxation that appears to be not significantly
perturbed by the brush potential. Therefore, we conclude that the
brush potential does not play an essential role for the relaxation
of the chains as postulated previously \cite{OConnor}. However, the
position vector auto-correlation data in direction perpendicular to
the grafting plane show a collective decay in contrast to the hierarchical
relaxation in parallel directions. This qualitatively different behavior
can be explained by the predominant orientation of the tube in perpendicular
direction which suppresses large scale reorientation of tube sections,
such that only fluctuations around an average orientation survive.
The brush potential turns the relaxation of these tube orientation
fluctuations into a time depending net pulling force acting on the
chains, which causes a collective motion of the chain monomers in
perpendicular direction after each relaxation step. These collective
modes are the dominating contribution in perpendicular direction beyond
the Rouse time of the chains, while the exponentially rare contribution
of deep retractions can be ignored. In parallel direction, however,
these collective modes enter only as a small correction to the classical
arm retraction dynamics.

At this point, we also would like to stress that hydrodynamic effects
are not accessible with our simulation method. Certainly, hydrodynamics
plays a significant role near the free end and may enhance the collective
motion of larger sections of a brush. However, arm retraction is the
process of changing the relative positions of two entangled strands,
which is difficult to achieve by a collective motion of the chains.
Furthermore, the grafting of the polymers damps out collective motion
inside the brush. Therefore, we do not expect a significant impact
of hydrodynamics for the scaling of the terminal relaxation time as
a function of $N$.

The results of the present work can be applied in further research
to understand the dynamics of e.g. bidisperse or polydisperse layers
of grafted chains. Polydispersity is most relevant for technical applications,
while bidisperse layers are often found in biological systems, for
instance, the brush made of MUC1 and MUC4 in the periciliary layer
\cite{Button}.

\section{Acknowledgments}

The authors thank the ZIH Dresden for a generous grant of computing
time, the DFG for funding projects LA 2735/2-1 and KR 2854/3-1, and
M. Hoffmann for providing the numerical solution of grafted Rouse
chains. T.C.B. McLeish, M. Rubinstein, H. Merlitz, and J.-U. Sommer
are thankfully acknowledged for helpful discussions.


\begin{thebibliography}{10}
\bibitem{footnote2}Note that there is a numerical coefficient connecting
the entanglement degree of polymerization and tube diameter as discussed
in \cite{Larson,Lang}.
\bibitem{footnote}For instance, if $\left\langle \left[d_{n}(\Delta t)\right]^{2}\right\rangle _{\perp}\propto t^{x}$
while $\left\langle \left[d_{n}(\Delta t)\right]^{2}\right\rangle _{\parallel}\propto t^{y}$,
we obtain $d_{r}\propto\left\langle \left[d_{n}(\Delta t)\right]^{2}\right\rangle _{\parallel}^{x/y-1}$.
\end{thebibliography}
\end{document}